\documentclass[twocolumn,aps,amsmath,amssymb,prl,superscriptaddress]{revtex4}
\usepackage{graphicx}
\usepackage{amsmath}
\usepackage{braket}

\begin{document}

\title{Time-resolved tomography of a driven adiabatic quantum simulation}

\author{Gian Salis}
\author{Nikolaj Moll}
\affiliation{IBM Research -- Zurich, S\"aumerstrasse 4, 8803 R\"uschlikon, Switzerland}
\author{Marco Roth}
\affiliation{Institute for Quantum Information, RWTH Aachen University, D-52056 Aachen, Germany}
\author{Marc Ganzhorn}
\author{Stefan Filipp}
\affiliation{IBM Research -- Zurich, S\"aumerstrasse 4, 8803 R\"uschlikon, Switzerland}

\date{January 14, 2020}

\begin{abstract}
A typical goal of a quantum simulation is to find the energy levels and eigenstates of a given  Hamiltonian. This can be realized by adiabatically varying the system control parameters to steer an initial eigenstate into the eigenstate of the target Hamiltonian. Such an adiabatic quantum simulation is demonstrated by directly implementing a controllable and smoothly varying Hamiltonian in the rotating frame of two superconducting qubits, including longitudinal and transverse fields and iSWAP-type two-qubit interactions. The evolution of each eigenstate is tracked using time-resolved state tomography. The energy gaps between instantaneous eigenstates are chosen such that depending on the energy transition rate either diabatic or adiabatic passages are observed in the measured energies and correlators. Errors in the obtained energy values induced by finite $T_1$ and $T_2$ times of the qubits are mitigated by extrapolation to short protocol times.
\end{abstract}


\maketitle

By the adiabatic theorem~\cite{Messiah1991}, a quantum system will remain in an initially prepared eigenstate when the parameters of the Hamiltonian are slowly modified. Such an adiabatic evolution will guide the system from a well-known and well-controlled eigenstate of a simple Hamiltonian $H_\textrm{i}$ to an a-priori unknown eigenstate of a complex target Hamiltonian $H_\textrm{t}$. This property can be exploited for adiabatic quantum computation~      \cite{farhiQuantumAdiabaticEvolution2001,lloydQuantumInformationMatters2008}, where  the interest is to find the ground state of $H_\textrm{t}$, typically an Ising-type Hamiltonian that may encode the solution of a given computational problem~\cite{Farhi2001}. The method is, however, not restricted to a specific application in computing. It is also well suited to study ground states of many-body quantum systems, such as spin
systems~\cite{King2018, kimQuantumSimulationFrustrated2010, roushanChiralGroundstateCurrents2017} or fermionic  matter~\cite{babbushAdiabaticQuantumSimulation2014,rothAdiabaticQuantumSimulations2019,Barkoutsos2017}. If there is no direct map between the Hamiltonian of the physical system and that of the quantum simulator~\cite{Bloch2012}, the system properties such as the type of particles and their interactions must be mapped onto the qubits first~\cite{Jordan1928,Bravyi2002}, and adiabatic parameter variations may not be possible on the simulator. To go beyond these limitations, the adiabatic protocol can be digitized~\cite{steffenExperimentalImplementationAdiabatic2003,barendsDigitizedAdiabaticQuantum2016,Moll2018,Farhi2014}, or extra ancillary degrees of freedom can be utilized~\cite{Kempe2006,Bravyi2008,Barkoutsos2017}. Alternatively, the adiabatic evolution of a quantum system can be tracked with a variational approach~\cite{Li2017,chenDemonstrationAdiabaticVariational2019}. This may provide an efficient way to generate problem-specific trial states~\cite{ganzhornGateEfficientSimulationMolecular2019} for the variational quantum eigensolver.

In either case, when scaling adiabatic quantum simulations to large systems, one has to keep in mind that the smallest energy gap may decrease with the number of qubits~\cite{canevaAdiabaticQuantumDynamics2008, albashTemperatureScalingLaw2017}, leading to an increase in the protocol time required to adiabatically pass the avoided crossing without inducing diabatic transitions to other energy eigenstates. The smallest energy gap depends on the specific parameters of of $H_\textrm{i}$ and $H_\textrm{t}$ and can be controlled by the path taken between the two. A possible way to increase the gap size is to switch on a ‘navigator’ Hamiltonian~\cite{Farhi2002} in form of extra, often transverse interaction terms in the middle of the process ~\cite{matsuuraVanQverVariationalAdiabatically2019,Nishimori2017}. Since in large systems the energy landscape and therefore the size and structure of the gaps during the evolution is not known, it is challenging to study their effect on the adiabatic dynamics directly and in-situ.

As we show here, this can be achieved by utilizing the toolbox from gate-based quantum computation and apply state tomography techniques to adiabatic protocols. We explore the adiabatic and diabatic passage of avoided crossings in a two-qubit system realized by parametrically coupled superconducting qubits~\cite{mckayEfficientGatesQuantum2017, rothAnalysisParametricallyDriven2017}. The Hamiltonian is implemented by smoothly controllable single-qubit fields and $XY$-type (iSWAP) interactions~\cite{kimQuantumSimulationTransverse2011}, enabling versatile analog adiabatic protocols. By preparing the system in either of its initial eigenstates, the evolution of energy and wave function of all excited states is tracked.  We study the passage of excited states through avoided level crossings as a function of gap size and transition time. 
By adjusting the interaction strength and the duration of the adiabatic protocol we control whether the passage occurs diabatically or adiabatically. 
Using state tomography based on single-shot readout of each qubit, we extract the energy eigenstates at each instant of time of the adiabatic protocol in excellent agreement with theoretical predictions. Noticeably, by evaluating the transient Pauli correlators of the wave function we can clearly differentiate adiabatic from diabatic passages.   
We then apply error mitigation by extrapolating to short operation times to lessen the effect of qubit dissipation, similar to the techniques appplied to digital circuits~\cite{temmeErrorMitigationShortDepth2017, liEfficientVariationalQuantum2017, kandalaErrorMitigationExtends2019}

\begin{figure}[tb]
\begin{center}
\includegraphics[width=\linewidth]{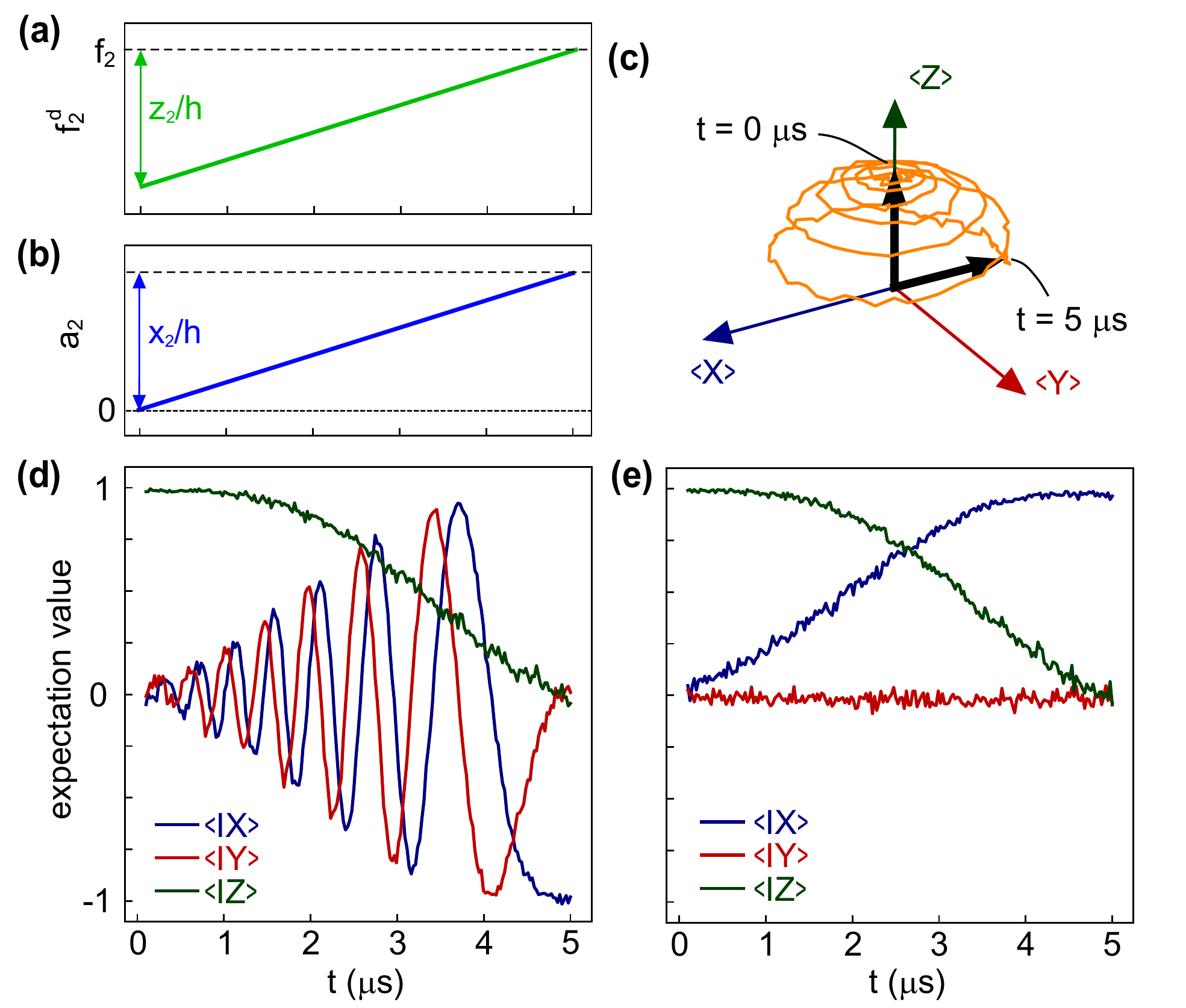}
\caption{\label{fig1} Adiabatic transition of qubit 2 from the $z$ into the $x$ direction: chirped adiabatic single qubit drive with (a) frequency offset $z_2/h$ providing an effective field along the $z$ direction that decreases with time, and (b) amplitude $x_2$ of the drive determining the field along $x$ that increases with time. (c) Measured evolution of qubits prepared at $t=0$ in the $\ket{00}$ state when exposed to chirped adiabatic single-qubit drive, as seen in the rotating frame at constant precession frequency $f_2$. (d) Measured expectation values $\braket{IX}$, $\braket{IY}$, and $\braket{IZ}$ of the qubit in the frame at constant precession frequency $f_2$, and (e) in a frame corrected for the accelerating single-qubit drive frequency.}
\end{center}
\end{figure}

Specifically, we realize the transition from $H_i=\frac{1}{2}(z_1ZI + z_2IZ)$ to $H_t=\frac{1}{2}(x_1 XI+x_2 IX)+\frac{1}{4}j(XX+YY)$ using two transmon-type qubits with fixed frequencies of $f_1^0 = 6.163~\rm{GHz}$ and $f_2^0 = 5.066~\rm{GHz}$. Here, $\{I,X,Y,Z\}$ denotes the set of Pauli operators, and we use short notation for the tensor product, e.g. $XI$ means $X \otimes I$ where $X$ acts on qubit 1 and $I$ on qubit 2. The qubits, which are fabricated on a silicon chip using standard Nb metallization and Al/Al$_2$O$_3$/Al Josephson junctions, are capacitively coupled to a common tunable coupler (TC) with a geometry similar to the one described in Ref.~\cite{mckayUniversalGateFixedFrequency2016}. The coupler is implemented as a flux-tunable transmon-type qubit with an asymmetric SQUID loop. Its  frequency can be varied between $4.82~\rm{GHz}$ and $8.26~\rm{GHz}$. For the measurements presented here, it is tuned to $7.85~\rm{GHz}$ by a dc flux-bias. An iSWAP-type gate is implemented by superposing an ac flux modulated at the difference frequency of the two involved qubits, which parametrically brings them into resonance~\cite{mckayUniversalGateFixedFrequency2016}.

Single-qubit fields along arbitrary directions are applied to each qubit $i$ in the frame rotating at the drive frequency $f^d_i$ by controlling the amplitudes $a_i(t)$, frequencies $f^d_i$ and phases $\phi_i^0$ of coherent microwave drives proportional to $a_i(t)\exp\{i\phi_i(t)t\}$ with $\phi_i(t) = \phi_i^0 + 2\pi (f_i - z_i/h)t + \pi z_i t^2/(ht_\textrm{ad})$. The Planck constant is denoted by $h$. This form of $\phi_i(t)$ linearly ramps the drive frequency $f^d_i = d\phi_i/dt$ from $f_i - z_i/h$ to $f_i$ within the adiabatic protocol time $t_\textrm{ad}$. In the frame rotating with $f^d_i$, qubit $i$ experiences an effective $Z$ interaction of size $\frac{1}{2}z_i(t_\textrm{ad}-t)/t_\textrm{ad}Z$  [Fig.~\ref{fig1}(a)]. The interaction of the transverse field is given by $\frac{1}{2}a_i(t) (\cos\phi_i^0 X + \sin\phi_i^0 Y)$~\cite{rothAdiabaticQuantumSimulations2019}. We linearly ramp up the $X$ interaction by setting $\phi_i^0=0$ and $a_i(t) = x_i t/t_\textrm{ad}$ [Fig.~\ref{fig1}(b)].

\begin{figure}[tb]
\begin{center}
\includegraphics[width=\linewidth]{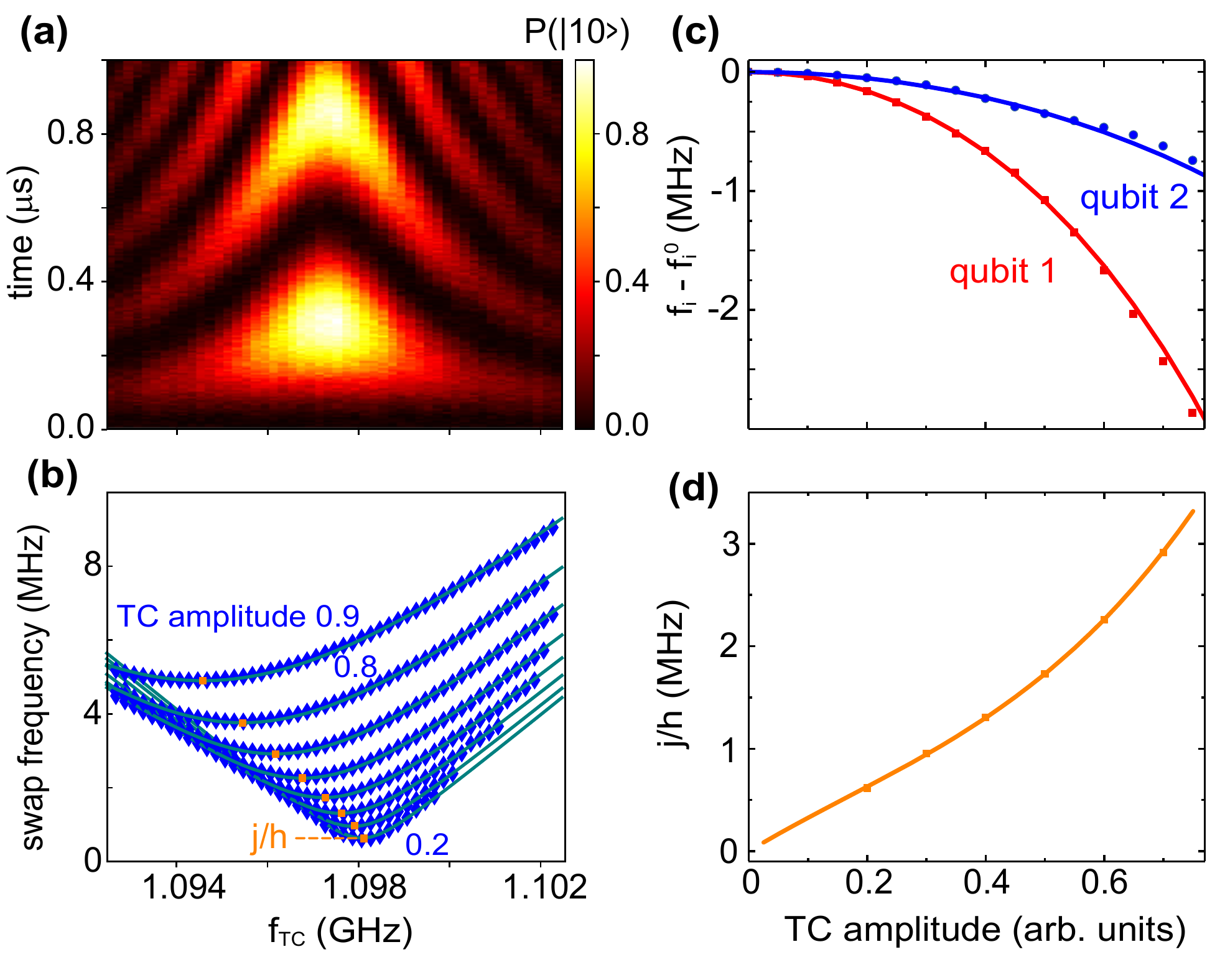}
\caption{\label{fig2} Calibration of iSWAP-type gate and dispersive qubit shifts. (a) Measured oscillations of the $\ket{10}$ state occupation after initialization in $\ket{01}$, for an ac flux modulation amplitude of $0.5$. (b) Fitted oscillation frequency of the $\ket{10}$ occupation (symbols) as a function of flux modulation frequency $f_\textrm{TC}$ for different modulation amplitudes from 0.2 to 0.9, and corresponding fit of its dependence on $f_\textrm{TC}$ (solid lines). (c) Dispersive shifts $f_i - f_i^0$ of individual qubits as induced by driving the iSWAP-type gate, measured using Ramsey spectroscopy of individual qubits in presence of an ac flux modulation. (d) Strength $j$ of iSWAP interaction as a function of drive amplitude, as obtained with $f_\textrm{TC}$ on resonance (minimum frequency in (b), orange squares).}
\end{center}
\end{figure}

We first discuss measurements where the iSWAP interaction is switched off. The two qubits are initialized in the $\ket{00}$ state such that the expectation value for the qubit polarizations $\braket{ZI}=\braket{IZ}=1$. Chirped microwave drives are then applied on both qubits with $z_i=3~\rm{MHz}$ and $x_i=2.7~\rm{MHz}$. Fig.~\ref{fig1}(c) and Fig.~\ref{fig1}(d) show the measured expectation values $\braket{IX}$, $\braket{IY}$ and $\braket{IZ}$ of the polarization of qubit 2 as a function of time, as seen in the reference frame rotating at the constant qubit frequency $f_2$. During the adiabatic protocol, the qubit polarization rotates about the $z$ axis and ends in the $x-y$ plane. In the reference frame adapted to the accelerating phase $\phi_i(t)$ of the microwave drive, the rotation disappears, see Fig.~\ref{fig1}(e), and an adiabatic transition of the qubit polarization from $z$ to $x$ direction is observed. Similar results are recorded for qubit 1 (not shown).

In a next step, we turn on the iSWAP-interaction between the qubits during the adiabatic protocol. Figure~\ref{fig2}(a) shows the characterization of the interaction, for which the qubits are initialized into the $\ket{01}$ state, and then an ac flux modulation at frequency $f_{\textrm{TC}}$ and constant amplitude is switched on for a time $t$. If $f_{\textrm{TC}}$ is close to $f_1^0 - f_2^0$, the gate swaps the $\ket{01}$ into the $\ket{10}$ state, and a measurement of the  occupation of qubit 1 displays the typical chevron-type pattern as a function of $t$ and $f_\textrm{TC}$. From such data, the frequency of the oscillation between the state $\ket{10}$ and $\ket{01}$ is determined for different flux modulation amplitudes and frequencies  $f_\textrm{TC}$ as shown in Fig.~\ref{fig2}(b). On resonance, a complete swap is realized, seen as a maximum in contrast and a minimum $j/h$ of the oscillation frequency (orange squares). As a consequence of the dispersive shifts of the two qubits, the resonance frequency decreases with increasing drive amplitude. The dispersively shifted qubit frequencies $f_i$ are determined independently using Ramsey spectroscopy while driving the TC 10\,MHz below the iSWAP resonance. The values of $f_i - f_i^0$ with a corresponding fit to a symmetric fourth-order polynomial are displayed in Fig.~\ref{fig2}(c). The shift of $f_2-f_1$ matches the shift of the resonance observed in Fig.~\ref{fig2}(b). The interaction strength $j$ first increases linearly with flux modulation amplitude and then superlinearly, see Fig.~\ref{fig2}(d). For the measurements described in this paper, the phase of the parametric drive is chosen such that the interaction is described by the term $\frac{j}{4}(XX+YY)$. By modifying this phase, additional $XY$ terms can be added~\cite{ganzhornGateEfficientSimulationMolecular2019}.

During the adiabatic protocol, the TC amplitude is linearly ramped from zero to a final amplitude, thereby increasing $j(t)$ monotonically, with the calibration shown in Fig.~\ref{fig2}(d). Throughout the protocol,  $f_{\textrm{TC}}$ is kept constant at the difference of the qubit frequency as reached at the end of the protocol. Also the linearly ramped single-qubit drive frequencies $f_i^d$ reach the dispersively shifted $f_i$ at the end of the protocol [Fig.~\ref{fig1}(a)]. The time-dependent Hamiltonian is then given by
\begin{multline}
\label{Eq1}
    H(t) = \left(1-\frac{t}{t_\textrm{ad}}\right)\left(z_1 ZI + z_2 IZ\right)/2 \\ + \frac{t}{t_\textrm{ad}}\left(x_1 XI + x_2 IX\right)/2 + j(t)(XX+YY)/4 \, .
\end{multline}

\begin{figure}[tb]
\begin{center}
\includegraphics[width=\linewidth]{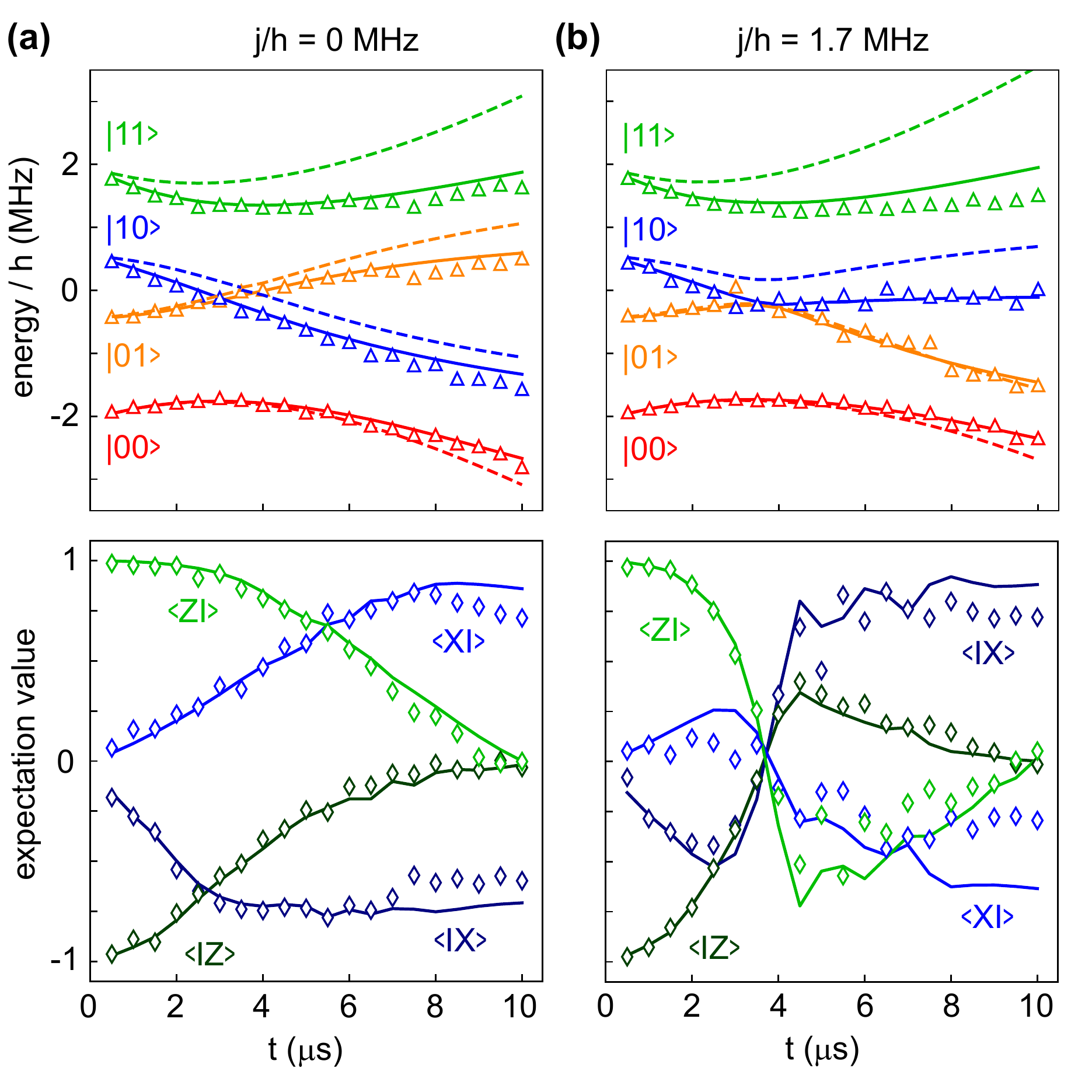}
\caption{\label{fig3} Time-resolved energy levels and Pauli-term expectation values during the adiabatic protocol for $z_1/h=2.5$\,MHz, $z_2/h=1.5$\,MHz, $x_1/h=2.0$\,MHz and $x_2/h=4.1$\,MHz. Because $z_1>z_2$ but $x_1<x_2$, the levels $E_{01}$ and $E_{10}$ cross during the protocol, with a gap that opens with increasing $j$. Symbols denote measured values, dashed lines exact energy values, and solid lines model calculations taking decoherence and thermalization into account. In (a) and (c), a diabatic passage is observed for $j/h=0$\,MHz, and in (b) and (d) an adiabatic passage for $j/h=1.7$\,MHz. In (c) and (d), the expectation values of the relevant Pauli terms are shown for starting in $\ket{01}$, with a smooth transition in (c) for diabatic passage, and an abrupt sign inversion at the adiabatic passage in (d).}
\end{center}
\end{figure}

To characterize the state at each instant of time, the state energy is determined by measuring the expectation values of each of the Pauli terms contained in $H(t)$, i.e.\ $\braket{ZI}$, $\braket{IZ}$, $\braket{XI}$, $\braket{IX}$, $\braket{XX}$, and $\braket{YY}$. These values are then multiplied with the respective coefficients of Eq.~(\ref{Eq1}) as set by the experiment, and summed over. In addition to initializing the qubits in the $\ket{00}$ state at the beginning of the protocol and thereby tracking the ground state energy, excited state energies $E_{01}$, $E_{10}$ and $E_{11}$ are obtained by initialization of the qubits in the $\ket{01}$, $\ket{10}$ and $\ket{11}$ states.

We choose single-qubit drives $z_1>z_2$ and $x_1<x_2$, leading to a crossing of $E_{10}$ and $E_{01}$ during the protocol if $j/h=0~\rm{MHz}$. The dashed lines in Fig.~\ref{fig3}(a) and (b) show the calculated energy levels of all four states during the adiabatic protocol. The 
experimentally obtained energy values are displayed as symbols. They initially match the calculated ones but increasingly deviate with progressing time, especially for higher excited states. This deviation can be accounted for by qubit decoherence and dissipation: a numerical calculation (solid lines) obtained by 
taking $T_1$ and $T_2$ into account through corresponding Lindblad operators using QUTIP~\cite{johanssonQuTiPOpensourcePython2012} shows excellent agreement with all measured eigenenergies. A small $ZZ$ interaction of size 200\,kHz originating from the indirect coupling of the two qubits via the TC is taken into account in the calculations.

For $j/h=0~\rm{MHz}$, the measured $E_{01}$ are below $E_{10}$ before the crossing, and above after it, corresponding to a diabatic passage of the crossing. For $j/h=1.7~\rm{MHz}$ [Fig.~\ref{fig3}(b)], a gap $a/h$ of size $0.38~\rm{MHz}$ opens. There, $E_{01}$ remains below $E_{10}$ throughout the protocol, evidence for an adiabatic passage of the avoided crossing. Due to qubit decoherence and dissipation, the gap is barely visible in the measured energies. The structure of the crossing becomes, however, clearly visible in the measured correlators. In Fig.~\ref{fig3}(c) and (d), the measured Pauli expectation values (correlators) $\braket{ZI}$, $\braket{IZ}$, $\braket{XI}$ and $\braket{IX}$ are shown for the $E_{01}$ state. For $j/h=0~\rm{MHz}$ in (c), the correlators $\braket{IZ}$ and $\braket{ZI}$ smoothly decrease to zero, whereas $\braket{IX}$ and $\braket{XI}$ increase towards positive and negative finite values. In strong contrast, for $j/h=1.7~\rm{MHz}$, all four correlators abruptly change at the avoided crossing. This clearly reflects the adjustment of the character of the wavefunction imposed by the adiabatic passage: at the avoided crossing, the $\ket{01}$-like wavefunction transitions into a  $\ket{10}$-like state, and vice versa, thereby reversing the signs of the corresponding Pauli correlators.

\begin{figure}[tb]
\begin{center}
\includegraphics[width=\linewidth]{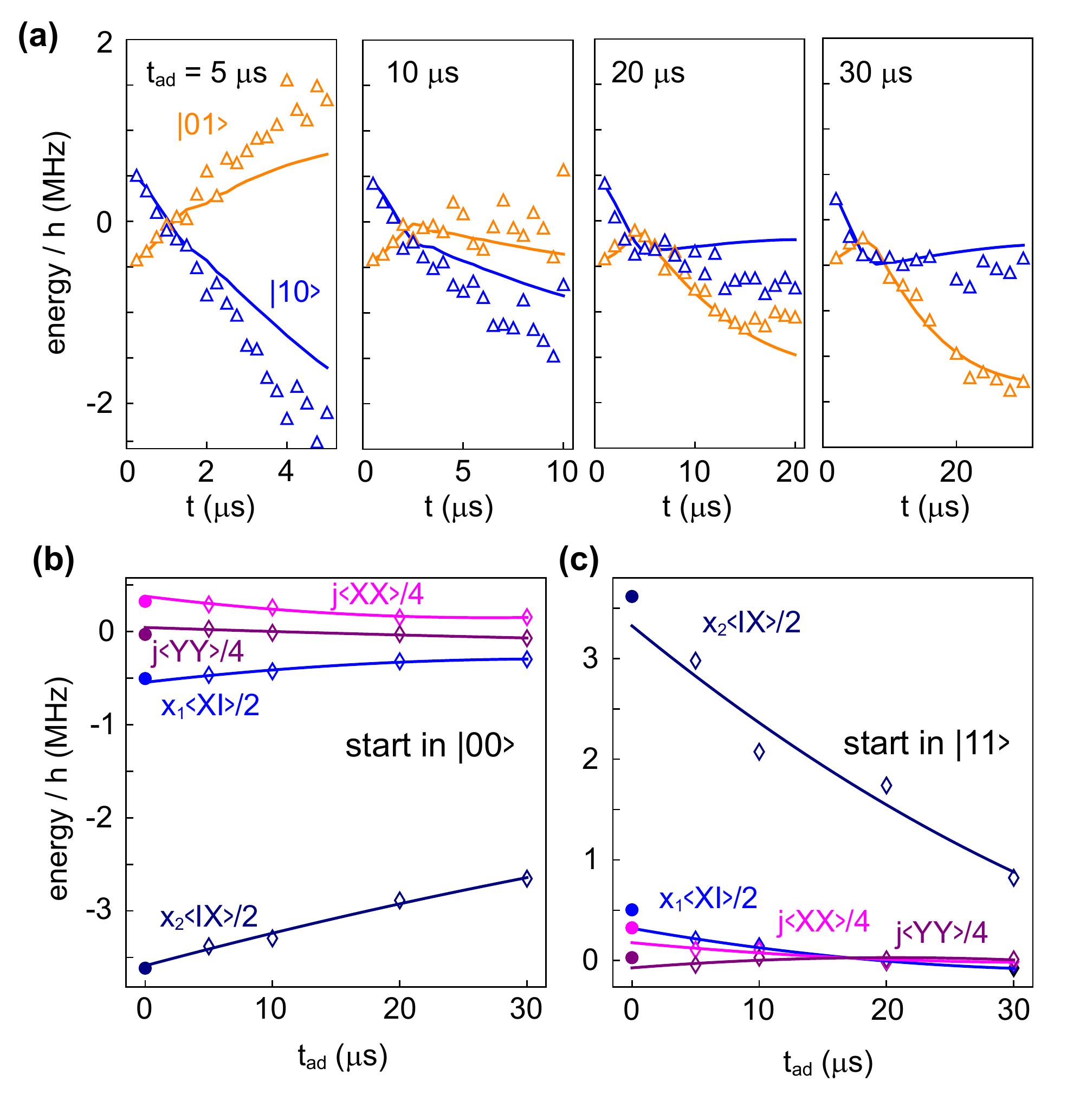}
\caption{\label{fig4} Influence of length $t_{\textrm{ad}}$ of the adiabatic protocol for $z_1/h=2.5~\rm{MHz}$, $z_2/h=1.5~\rm{MHz}$, $x_1/h=1.0~\rm{MHz}$, $x_2/h=7.3~\rm{MHz}$ and $j/h=1.3~\rm{MHz}$. (a) Cross-over from diabatic to adiabatic passage. Shown are energy levels $E_{01}$ and $E_{10}$ as a function of $t$, with a transition from a diabatic passage for $t_{\textrm{ad}} = 5~\mu\rm{s}$, and an adiabatic passage for $t_{\textrm{ad}}=30~\mu\rm{s}$. Symbols are measurements, solid lines  calculations. Longer times $t_{\textrm{ad}}$ lead to a stronger decay of the measured expectation values of the Pauli terms, reflected as an overall decrease of the absolute energy values. Individual energy contributions $x_1/2\braket{XI}$, $x_2/2\braket{IX}$, $j/4\braket{XX}$ and $j/4\braket{YY}$ are shown when starting in (b) the ground state $\ket{00}$ and (c) in the highest excited state $\ket{11}$. The energy contributions decay with longer $t_{\textrm{ad}}$, which is more pronounced for $\ket{11}$ than for $\ket{00}$. Open diamonds: measured values, solid lines: fitted second-order polynomials extrapolated to $t_\textrm{ad}=0$, filled dots at $t_\textrm{ad}=0$: calculated energy contributions.}
\end{center}
\end{figure}

Next we study the influence of the transition rate on the passage of the avoided crossing: the state with an initially lower energy ends either in the upper state (diabatic passage) or stays in the lower state (adiabatic passage), depending on the relation of the minimum gap $a$ with the transition rate $\alpha = \frac{\partial}{\partial{t}}(E'_1-E'_0)$, where $E'_0$ and $E'_1$ denote the linearly varying energies of the diabatic states. According to the Landau-Zener theory~\cite{wittigLandauZenerFormula2005}, the probability of a diabatic transition is given by $\exp({-2\pi\Gamma})$ with $\Gamma = a^2/(4\hbar|\alpha|)$. In the measurement presented in Fig.~\ref{fig4}(a), we have set a minimum gap $a/h$ of $0.22~\rm{MHz}$ and run the same adiabatic protocol for different protocol times $t_\textrm{ad}$. For $t_\textrm{ad}=5~\mu\rm{s}$, we find a diabatic passage. This can be continuously tuned to an adiabatic passage by increasing $t_\textrm{ad}$, see measurements for $t_\textrm{ad} = 10,\ 20$ and $30~\mu\rm{s}$. The transition rate $\alpha/h$ in this experiment is given by $10.3~\rm{MHz}/t_\textrm{ad}$, which gives a probability of $0.5$ for a diabatic transition at $t_\textrm{ad} = 15~\mu\rm{s}$,  in good agreement with the experimental data. The numerically calculated energies reproduce the time-resolved experimental energies very well. 

As mentioned above, the finite qubit lifetimes lead to a modification of the measured energy. This effect is more pronounced for the highest excited state, $\ket{11}$ than for the ground state $\ket{00}$. A mitigation is possible by extrapolating to small adiabatic protocol times. In Fig.~\ref{fig4}(b) and (c) we show the four contributions to the state energy, $\frac{1}{2}x_1\braket{XI}$, $\frac{1}{2}x_2\braket{IX}$, $\frac{1}{4}j\braket{XX}$ and $\frac{1}{4}j\braket{YY}$, for both the ground state energy $E_{00}$ and the highest excited state energy $E_{11}$. For the selected parameters, the largest contribution comes from $\braket{IX}$, followed by $\braket{XI}$ and $\braket{XX}$. The decay of these terms with increasing $t_\textrm{ad}$ as seen in Fig.~\ref{fig4}(b) can be mitigated by extrapolating the data available for the various $t_\textrm{ad}$ to $t_\textrm{ad}=0$, in a similar way as has been proposed for circuit-based algorithms in \cite{temmeErrorMitigationShortDepth2017}. Here, the dependencies of the individual Pauli terms on $t_\textrm{ad}$ are fitted by second-order polynomials. The values extrapolated to $t_\textrm{ad}=0$ are much closer to the exact energy contributions (shown as dots) than the values measured at $t_\textrm{ad}=5\mu$s, see table~\ref{tab1}. This type of error mitigation requires more attention if the shorter protocol times lead to diabatic passages of energy gaps. In that case, the order of the energy levels changes, as seen in Fig~\ref{fig4}(a). In this respect, a combination with adiabatic protocols starting with higher excited states ('going lower by aiming higher'~\cite{crossonDifferentStrategiesOptimization2014}) seems promising. 

\begin{table}
\setlength{\tabcolsep}{10pt}
\caption{\label{tab1} Energy values (in MHz) obtained by adiabatic protocols and extrapolation to zero protocol time.}
\begin{tabular}{ c  c  c  c  c }
\hline
\hline
  energy level & $t_\textrm{ad}=5\mu$s & extrapolation  & exact\\ \hline	
  $E_{00}/h$ & -3.52 & -3.71 & -3.82 \\
  $E_{11}/h$ & 3.25 & 3.75 & 4.48 \\
\hline
\hline
\end{tabular}

\end{table}

In conclusions, we have tested the performance of a driven two-qubit test system for adiabatic quantum simulations. Combining a smooth variation of the Hamiltonian parameters with digital circuit-based state preparation and two-qubit state tomography, we have obtained time-resolved energy values of ground state and excited states during the adiabatic evolution. We have studied the passage through avoided crossings with energy gaps and transition rates set by experimental parameters. Errors in the measured state energies can be mitigated by extrapolation to short protocol times. Scaling to more than two qubits could be achieved by connecting more qubits to a tunable coupler~\cite{rothAdiabaticQuantumSimulations2019} or by realizing extra ancilla degrees of freedom~\citep{Kempe2006}. We have restricted this study to an iSWAP-type two-qubit interaction term given by $XX+YY$, which can be used to model a system of exchange-coupled spins to test e.g. phase transitions in a Lipkin-Meshkov-Glick
model~\cite{canevaAdiabaticQuantumDynamics2008}. More generally, individual control of $XX$, $YY$, $XY/YX$ and even $ZZ$ contributions are possible in principle using the same quantum hardware by multichromatic modulation of the tunable coupler frequency~\cite{rothAdiabaticQuantumSimulations2019} and by controlling the phase of TC modulation~\cite{ganzhornGateEfficientSimulationMolecular2019}. This may open the door to solve more complex quantum problems~\cite{babbushAdiabaticQuantumSimulation2014} using adiabatic protocols.

Support from the quantum team at IBM T.~J.~Watson Research Center as well as valuable discussions with Matthias Mergenthaler, Andreas Fuhrer, Clemens M\"uller, Daniel Egger, Marek Pechal, Max Werninghaus, Stefan Paredes and Ivano Tavernelli are acknowledged. This work was supported by the IARPA LogiQ program under contract W911NF-16-1-0114-FE and by the EU Quromorphic project (grant No. 828826).

%\bibliographystyle{apsrev4}
%\bibliography{salis,IBMRefDB}

\end{document}